\documentclass[prl,noshowkeys,twocolumn]{revtex4}
\usepackage{graphicx}% Include figure files
\usepackage{dcolumn}% Align table columns on decimal point
\usepackage{bm}% bold math
\usepackage{color}
\usepackage[hidelinks]{hyperref}
\usepackage{amssymb,amsfonts,amsmath}

\newcommand{\rem}[1]{}

\begin{document}
\title{Colloidal trains}
\author{Mahla Mirzaee-Kakhki$^1$}
\author{Adrian Ernst$^{1}$}
\author{Daniel de las Heras$^{2}$}
\author{Maciej Urbaniak$^3$}
\author{Feliks Stobiecki$^3$} 
\author{Andreea Tomita$^4$}
\author{Rico Huhnstock$^4$}
\author{Iris Koch$^4$}
\author{Jendrik G\"ordes$^4$}
\author{Arno Ehresmann$^4$}
\author{Dennis Holzinger$^4$}
\author{Meike Reginka$^4$} 
\author{Thomas M. Fischer$^{1}$}
\email{Thomas.Fischer@uni-bayreuth.de}
\affiliation{
$^1$ Experimentalphysik X,  Physikalisches Institut, Universit{\"a}t Bayreuth, D-95440 Bayreuth, Germany.\\
$^2$Theoretische Physik II, Physikalisches Institut, Universit{\"a}t Bayreuth, D-95440 Bayreuth, Germany.\\
$^3$Institute of Molecular Physics, Polish Academy of Sciences, 60-179 Pozna\'n, Poland.\\
$^4$Institute of Physics and Center for Interdisciplinary Nanostructure Science and Technology (CINSaT), University of Kassel, D-34132 Kassel, Germany}
\date{\today}

\begin{abstract}
Single and double paramagnetic colloidal particles are placed above a magnetic square
pattern and are driven with an external magnetic field precessing around a high
symmetry direction of the pattern. The external magnetic field and that of the pattern
confine the colloids into lanes parallel to a lattice vector of the pattern.
The precession of the external field causes traveling minima of the magnetic potential along
the direction of the lanes. At sufficiently high frequencies of modulation only the doublets respond to the external field and move in direction of the traveling minima along the lanes, while the single colloids
cannot follow and remain static. We show how the doublets can
induce a coordinated motion of the single colloids building colloidal trains made of a chain
of several single colloids transported by doublets.

\end{abstract}
% insert suggested PACS numbers in braces on next line
%\pacs{82.70.Dd 	Colloids, 87.15.hm 	Folding dynamics, 87.19.St 	Movement and locomotion}
\maketitle
Biomimetics is used to implement biological functions to artificial devices, fulfilling tasks in a
non biological environment. Well known examples are artificial swimmers~\cite{Bibette,Snezhko,Martin}
and active systems \cite{active} that can be used to e.g. transport a load. Microscopic dynamics can also be inspired by large scale transport systems such as trains.
A railroad train is powered either by a separate locomotive or by multiple units of self propelled
equally powered carriages. In nature the motility of family members of animals trailing behind
each other is neither concentrated in the animal heading the trail nor is it distributed equally
amongst family members. Young goslings trailing behind one of their parents need less but not
zero power to follow their more powerful mother goose \cite{Ducklings}. When elephants travel
they walk in a line placing their youngest in between the grownups with a grownup at the head
and at the tail. In the spirit of other bioinspired magnetic colloidal
dynamics~\cite{Snezhko,Martin,Bibette,ribbon,aster} we generate a biomimetic train of a
collective ensemble of paramagnetic colloids. Single colloids are too weak to move on
their own along the line and must be assisted to move by
%hydrodynamically
pushing of the train with a larger paramagnetic colloidal doublet.  The train is confined to an
effectively one dimensional lane created via a colloidal potential which is generated by
the combination of a magnetic square pattern and an external magnetic field. The power of
each unit in the train is generated by modulating
the external field on a control loop. Colloidal trains can only move above a square lattice if
a sufficiently flat potential valley is created by orienting the external field roughly in direction
of a primitive unit vector of the square magnetic pattern.

\begin{figure*}
\includegraphics[width=1.9\columnwidth]{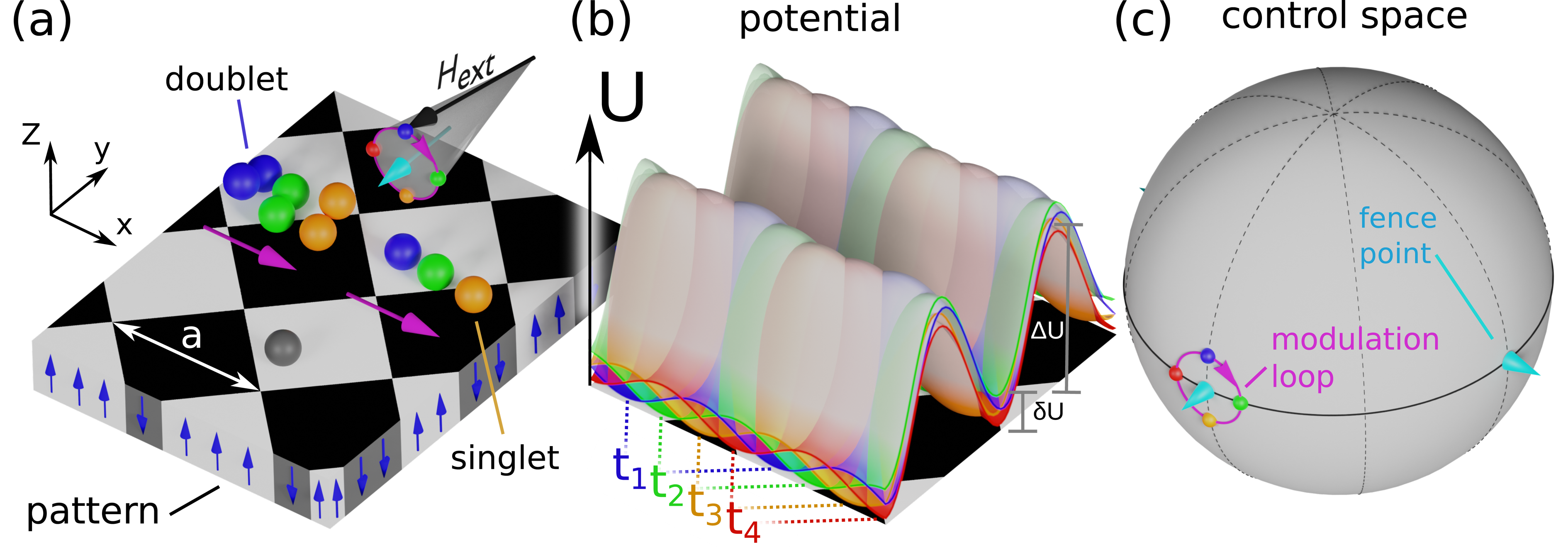}
\caption{Schematic of the setup. (a) Single and double spherical colloidal particles are placed on top of a
lithographic magnetic square pattern (lattice constant $a\approx7$ $\mu{\rm m}$) of up (white) and down (black) magnetized domains along
the ${\bf z}-$axis. An isolated single colloid (gray) does not move at the modulation frequency
applied in the experiments. Doublets are able to respond to the external field and can also induce
motion of single colloids. The color (blue, green, and orange) of the doublet and of the singlet
represents the position at three different times ($t_1$,$t_2$, and $t_3$) during the modulation loop.
(b)  Magnetic potential $U$ created by the pattern and the external potential
for four different times ($t_i$, $i=1,2,3,4$) during one modulation loop (red,blue,green, and orange) of the external field. The ratio between the
large and the small barriers of the potential is $\Delta U/\delta U\approx10$.
(c) Each point in control space $\cal C$ (gray sphere) corresponds to a different 
orientation of the external field. The experimental modulation loop  is highlighted in purple.
The loop winds around a fence point (cyan) of control space. The colors
of the four potentials in (b) and the coloring of the moving colloidal positions (a) correspond
to the four colored points on the loop in control space.
}
\label{fig1}
\end{figure*}

We illustrate the doublet assisted motion of a train of single colloids using a square
magnetic lattice \cite{tp2,tp3}, Fig.~\ref{fig1}(a). In the experiments, single paramagnetic
colloidal particles or doublets of particles move on a plane above a thin Co/Au layered system with
perpendicular magnetic anisotropy lithographically patterned via ion
bombardment~\cite{CBF1998,KET2010,tp3}. The pattern is a square lattice of magnetized
domains with a mesoscopic pattern lattice constant $a\approx 7\,\mu\rm{m}$, see a sketch
in Fig.~\ref{fig1}(a). The pattern is magnetized in the $\pm {\bf z}$-direction, normal to
the film. The pattern is spin coated with a $1.6\, \mu m$ polymer film that serves
as a spacer between pattern and the colloids.
The paramagnetic colloidal particles (diameter $2.8\,\mu\rm{m}$) are
immersed in water.  A uniform time-dependent external magnetic field $\mathbf{H}_{\rm{ext}}$
of constant magnitude ($H_{\rm{ext}}=4\,\rm{kAm}^{-1}$) is superimposed to the non-uniform
and time-independent magnetic field generated by the pattern $\mathbf{H}_p$ ($H_p\approx 1\,\rm{kAm}^{-1}$).
The external field is strong enough such that some of the paramagnetic particles self-assembly
into doublets due to induced dipolar interactions between the single colloidal particles. 
The doublets then align with the direction of the external field.
Our control space $\cal C$ is the surface of a sphere that represents all possible orientations
of the external field. We vary the external field $\mathbf{H}_{\rm{ext}}(t)$ with time $t$ performing 
periodic closed modulation loops in $\cal C$.

%but avoiding the directions
%of the external field along primitive unit vectors. The topology of our control space $\cal C$ is
%that of a punctured sphere. Fence points (cyan) along the primitive unit vector directions are
%removed from the sphere which renders the sphere not simply connected, see Fig.  \ref{fig1}c.
%We perform periodic modulation loops $\cal L_C$ of the external field in control space $\cal C$ to drive the system.
%We call the plane where the colloids move the action space $\cal A$. 

Both $\mathbf{H}_p$ and $\mathbf{H}_{\rm{ext}}$ create a potential $U\propto-\mathbf{H}_{\rm{ext}}\cdot\mathbf{H}_p$.
The potential is a periodic function of the position of the colloids in the magnetic lattice and it depends parametrically
on the orientation of the external field in control space $\cal C$.
At every time during the modulation loop, the colloids are attracted toward the minima of the potential. 
Full details about the computation of $U$ and the motion of single colloids are given in Refs.~\cite{tp2,tp3}. Here, we
briefly explain the points relevant for the present study. For a square lattice, there exist four special points (fence points) in
$\cal C$. The four points represent four directions of $\mathbf{H}_{\rm{ext}}$ which are parallel and antiparallel to
the lattice vectors of the square pattern. If the modulation loop of $\mathbf{H}_{\rm{ext}}$ in $\cal C$ winds
around one of the fence points, then the minima of the potential move one unit cell above the square pattern~\cite{tp2,tp3}.
The motion is topologically protected, with the set of winding numbers around the fence points defining the topological invariants.

\begin{figure}
\includegraphics[width=.90\columnwidth]{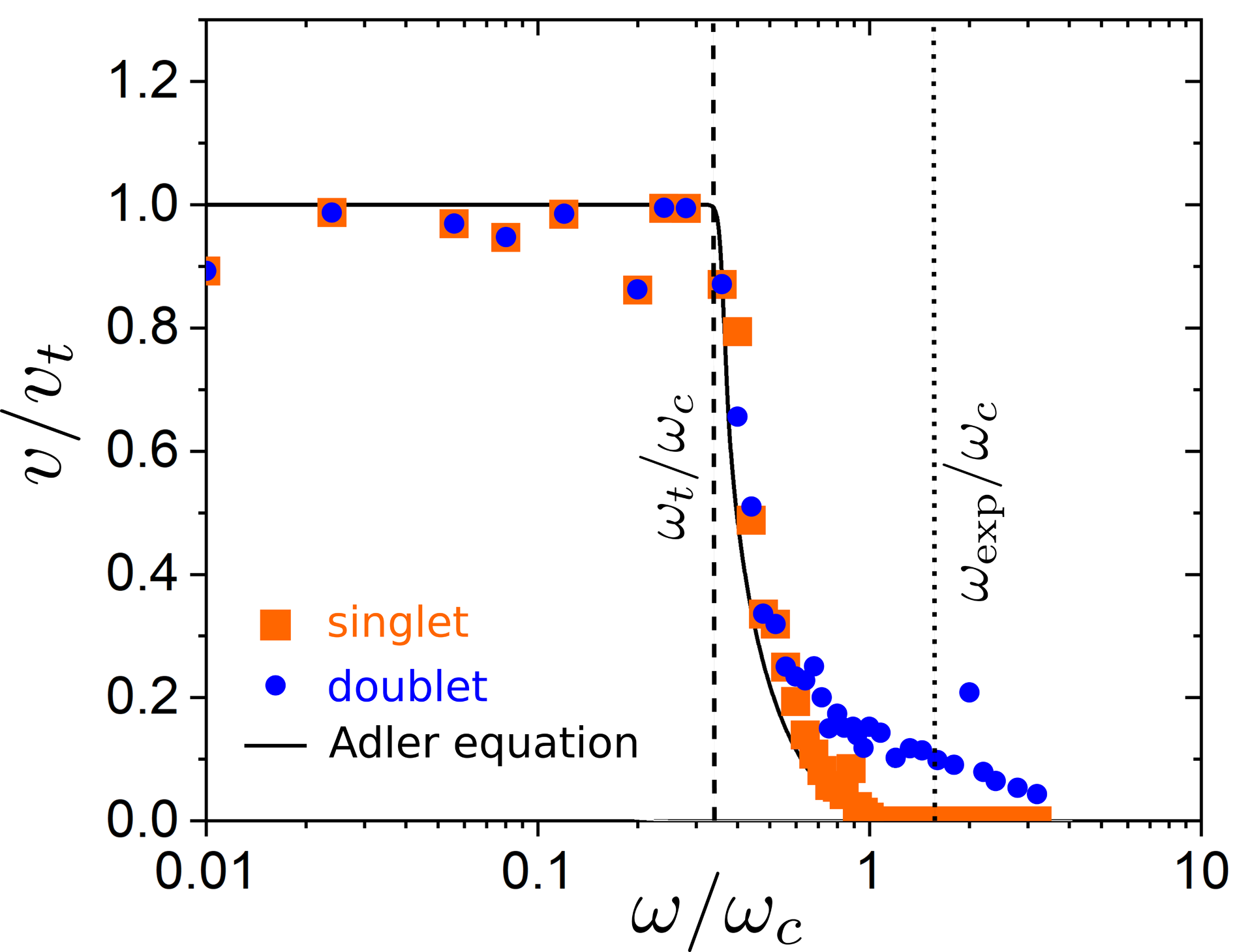}
\caption{Scaled magnitude of the velocity of singlets (orange squares) and doublets (blue circles) as 
a function of the scaled driving frequency. The black solid line is a fit of the 
singlet velocity using the generalized Adler equation~\eqref{EQadler}. The vertical
lines indicate the scaled experimental frequency $\omega_{\rm ext}$ and the topological
frequency $\omega_t$ below which the motion is topologically protected.
%The ratio between the singlet (doublet ) velocity and the topological velocity as a function
%of the driving frequency. The fit is performed with a generalized Adler equation (1). The angular
%frequency of the experiments must be chosen above the singlet stalling angular frequency
%$\omega_{exp}>\Omega_c$.}
}
\label{fig2}
\end{figure}

If $\mathbf{H}_{\rm{ext}}$ points in the direction of a fence point, the magnetic potential $U$
is effectively one-dimensional with valleys along the direction perpendicular to $\mathbf{H}_{\rm{ext}}$.
For example, let the lattice vectors of the magnetic pattern point along $\mathbf{x}$ and $\mathbf{y}$,
Fig.~\ref{fig1}(a). The magnetic potential $U$ exhibits deep valleys along $\mathbf{x}$, Fig.~\ref{fig1}(b),
when $\mathbf{H}_{\rm{ext}}$ points along $-\mathbf{y}$, Fig.~\ref{fig1}(c). If $\mathbf{H}_{\rm{ext}}$
slightly deviates from the direction of a fence point, see modulation loop in Fig.~\ref{fig1}(c), then
secondary minima of $U$ appear along the direction of the valleys, Fig.~\ref{fig1}(b).
The variation of the potential along the valley $\delta U$ is much smaller than in the transversal direction $\Delta U$.
In our experimental setup, we find $\Delta U /\delta U\approx 10$. 
%In Fig.  \ref{fig1}b we plot the potential $U$ for four directions
%(red,blue,green,orange) of the external field. For an external field pointing roughly into the $-y$-direction
%the potential exhibits deep valleys along the x direction, that confine the colloids in such a valley.
%We call the avoided external field directions along the primitive unit vectors of the lattice the fence points.
Modulating the external field directions along a loop that encloses a fence point, Fig.~\ref{fig1}(c), causes the
minima of $U$ to travel by one unit vector of the lattice upon completion of the loop. 
%The colors of the
%four potentials in Fig.  \ref{fig1}(b) and the coloring of the moving colloidal positions (Fig.~\ref{fig1}(a)
%correspond to the four colored points on the loop in control space Fig.  \ref{fig1}c.
The frequency of the loop $\omega$ can be chosen such that single colloids cannot follow the potential
minimum on their own but doublets move in direction of the traveling minima. If a doublet is on a collision course with a single colloid, then
the doublet can render the single colloid mobile and drive it through the potential valley, Fig.~\ref{fig1}(a).

\begin{figure*}
\includegraphics[width=2.00\columnwidth]{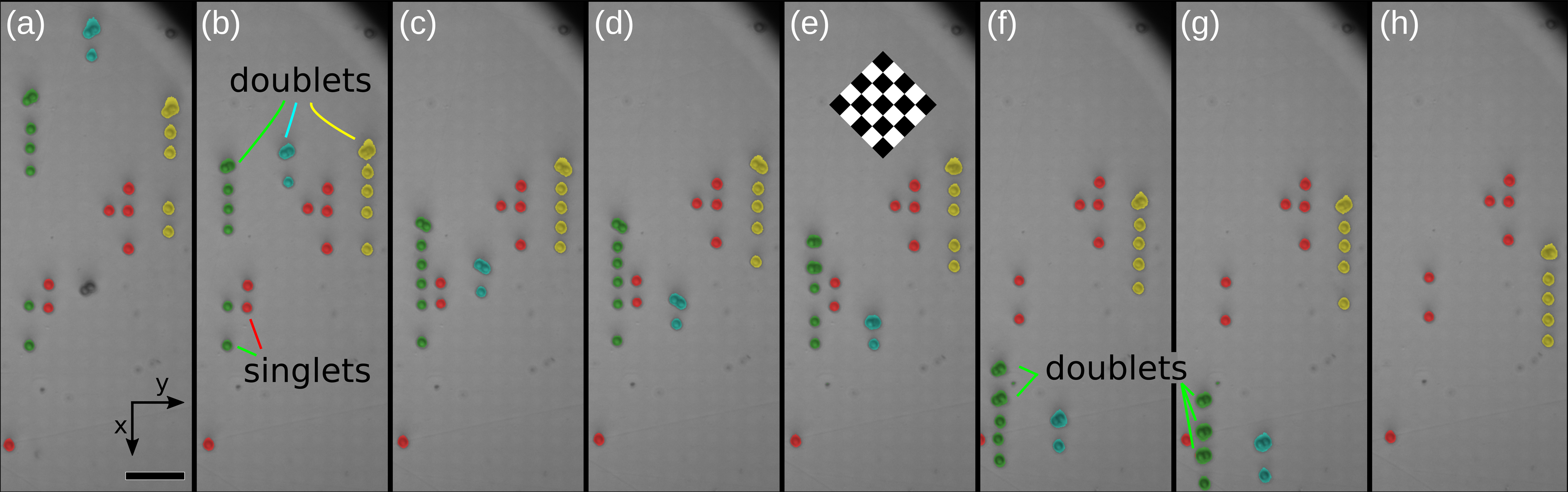}
\caption{Time sequence of microscope images of the pattern illustrating the motion of colloidal trains. Each train
has been artificially colored differently.
The images correspond to the times (a) $t=0$ s, (b) $t=8$ s , (c)
$t=13$ s (d) $t=15$ s (e) $t=17$ s (f) $t=23$ s (g) $t=25$ s and (h) $t=30$ s.
The period of one modulation loop is $2\pi/\omega_{\rm exp}\approx0.12$ s.
Scale bar is $20$ $\mu$m. A sketch of the pattern has been superimposed in (e).
A video clip recorded at twenty frames per second showing the motion of trains
and the non motion of singlets is provided in the supplemental material (adfig3a.mp4).
A second video clip (adfig3b.mp4) recorded at sixty frames per second shows a time-resolved
slow motion of the doublet during the course of a few modulation loops.}
\label{fig3}
\end{figure*}

In Fig.~\ref{fig2} we plot the speed of an isolated single colloid and that of an isolated
doublet versus the driving angular frequency $\omega$ of the modulation loop in $\cal C$.
At low frequencies, lower than the topological critical frequency $\omega_t$,
the motion is adiabatic and topologically protected. Singlets and doublets follow the potential
minimum at all times. Hence, the displacement caused by a loop is topologically locked to the primitive
lattice vector
%$a\mathbf e_x$ 
and particles move at the topological speed given by the lattice constant $a$ times the
frequency, i.e., $v_{t}=\omega a/2\pi$. However, at higher frequencies $\omega>\omega_t$ the
speed drops below $v_t$, mostly due to viscous and adhesive forces impeding the motion. For singlets, the speed decreases
with increasing frequency until the critical frequency $\omega_c$ is reached. At $\omega_c$ the isolated single colloids stop moving.
The speed of the singlets is well described by a generalized Adler equation~\cite{Adler}
\begin{equation}
v/v_{t}=\left\{\begin{array}{lcc}
1&&\rm{ if }\;\omega\le\omega_t\cr 
1-\sqrt{\frac{(\omega-\omega_t)((\omega_c-\omega)\omega_t +\omega\omega_c)}{(\omega_c-\omega_t)\omega^2}}&&\rm{if}\;\omega_t<\omega<\omega_c\cr
0&&\rm{if}\;\omega\ge\omega_c
\end{array}\right.\label{EQadler}
\end{equation}
The force due to the potential acting on a doublet is roughly twice the force acting on a single
colloidal particle. The viscous friction on the doublet, however, is less than twice the friction of a
single colloid because of hydrodynamic interactions  \cite{Happel}.
Hence, the doublet can still move at frequencies higher than the critical frequency of the singlets,
and we have a regime where the doublet moves while the singlet is at rest.
The experiments are performed at an angular frequency of $\omega\approx52\;s^{-1}>\omega_c$, such that singlets do not move, and doublets
move with a speed of roughly one eighth of the topological speed ($v_{\rm d}/v_t\approx0.125$), see Fig.~\ref{fig2}.

\begin{figure}
\includegraphics[width=1.\columnwidth]{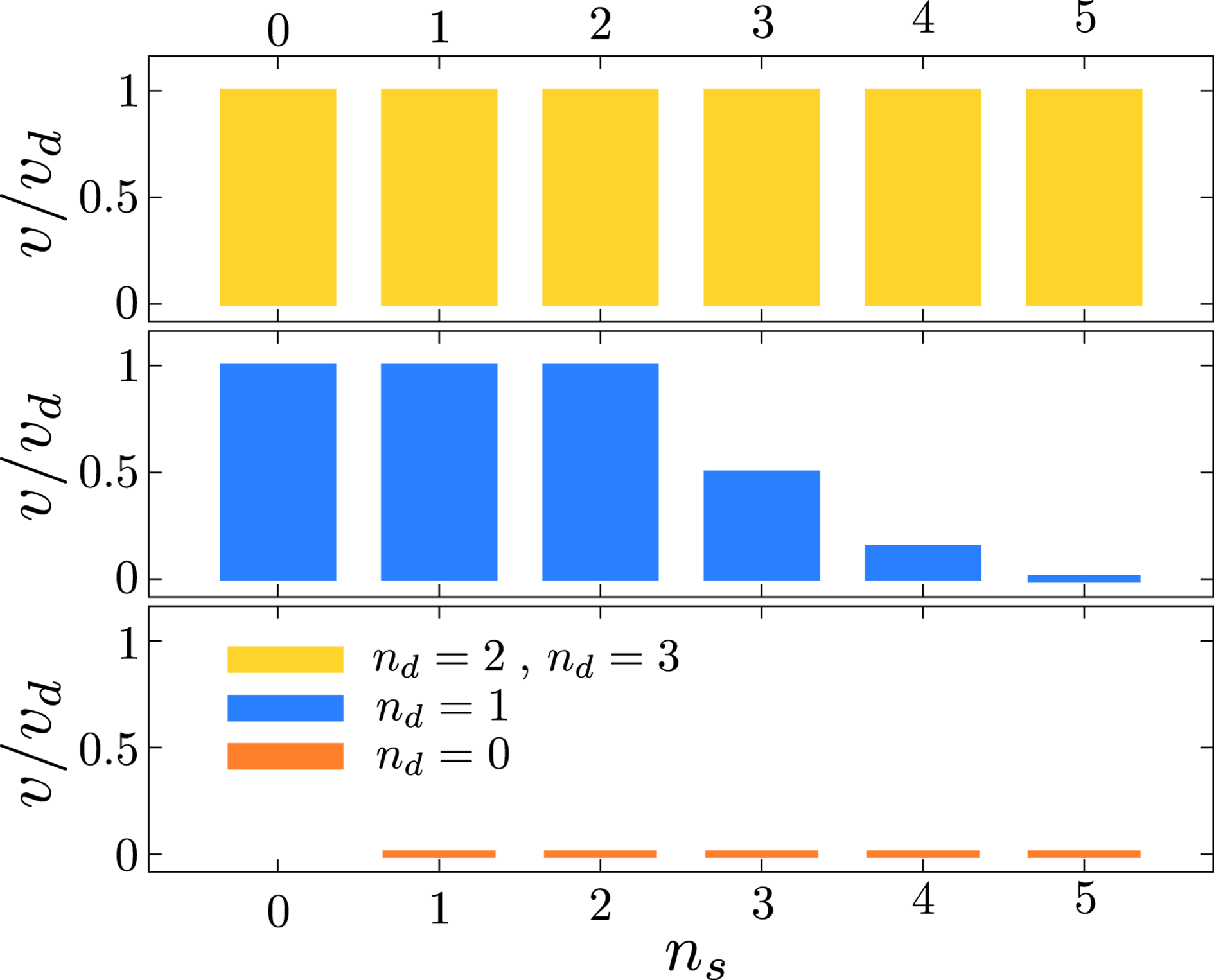}
	\caption{Speed of the train (scaled with the speed of a doublet $v_d$) versus the load, i.e. the number of singlets in the train $n_s$.
	 Data for different numbers of doublets in front of the train: $n_d=0$ (orange), $n_d=1$ (blue), $n_d=2$ and $3$ (yellow).}
\label{fig4}
\end{figure}

In Fig.~\ref{fig3} we show a time sequence of microscope images showing the dynamics of the colloids.
We distinguish single colloids (red) that are immobile from colloidal trains (green, cyan and yellow)
consisting of one, two, and three doublets at the tail, and of one up to five single colloids
at the front of the trains. The singlets and doublets in a train are well separated from each other by a primitive unite vector of the lattice. All trains move into the positive $\mathbf{x}$-direction with the doublets pushing the
singlets. Nothing particular happens to the cyan train with one doublet and one singlet moving through
the field of view at the doublet speed $v_{d}$. The green train with one doublet and three
singlets moves with half the doublet speed, Fig.~\ref{fig3}(a-b), collects two further singlets, Fig.~\ref{fig3}(c),
and stalls, Fig.~\ref{fig3}(d), until the two singlets close to the pushing doublets form a second doublet, Fig.~\ref{fig3}(e).
This increases the power of the train such that it resumes to move, Fig.~\ref{fig3}(e-f) at the doublet
speed. In Fig.~\ref{fig3}(g) a third doublet is formed leaving only one singlet in the green train before it
exits the field of view. Interestingly, when the green train passes the red singlets (sitting on the next track
to the right) the front immobile red single colloid is mobilized and performs
a single translation by one unit vector in the positive $\mathbf{x}$-direction, compare the position of
the red colloids next to the green train in Figs.~\ref{fig3}(e) and~\ref{fig3}(f).
The yellow train originally consists of one
doublet and two singlets, Fig.~\ref{fig3}(a). It collects two further singlets, Figs.~\ref{fig3}(b,c), and then moves as one
doublet and four singlets train at a relatively slow speed $v\approx0.15v_d$ through the image. No further
doublets are formed from the singlets of the yellow train as it moves. 
In Fig.~\ref{fig4} we plot the speed of a train as a function of both the number of singlets in front of the train and 
the number of doublets at its tail. A train with no doublet is immobile and a train with more than one
doublet can push up to five single colloids at the unloaded doublet speed. For one doublet we see a gradual
transition from motion at the doublet speed $v_d$ for trains with up to two singlets toward no motion for trains with five singlets.

\begin{figure}
\includegraphics[width=.9\columnwidth]{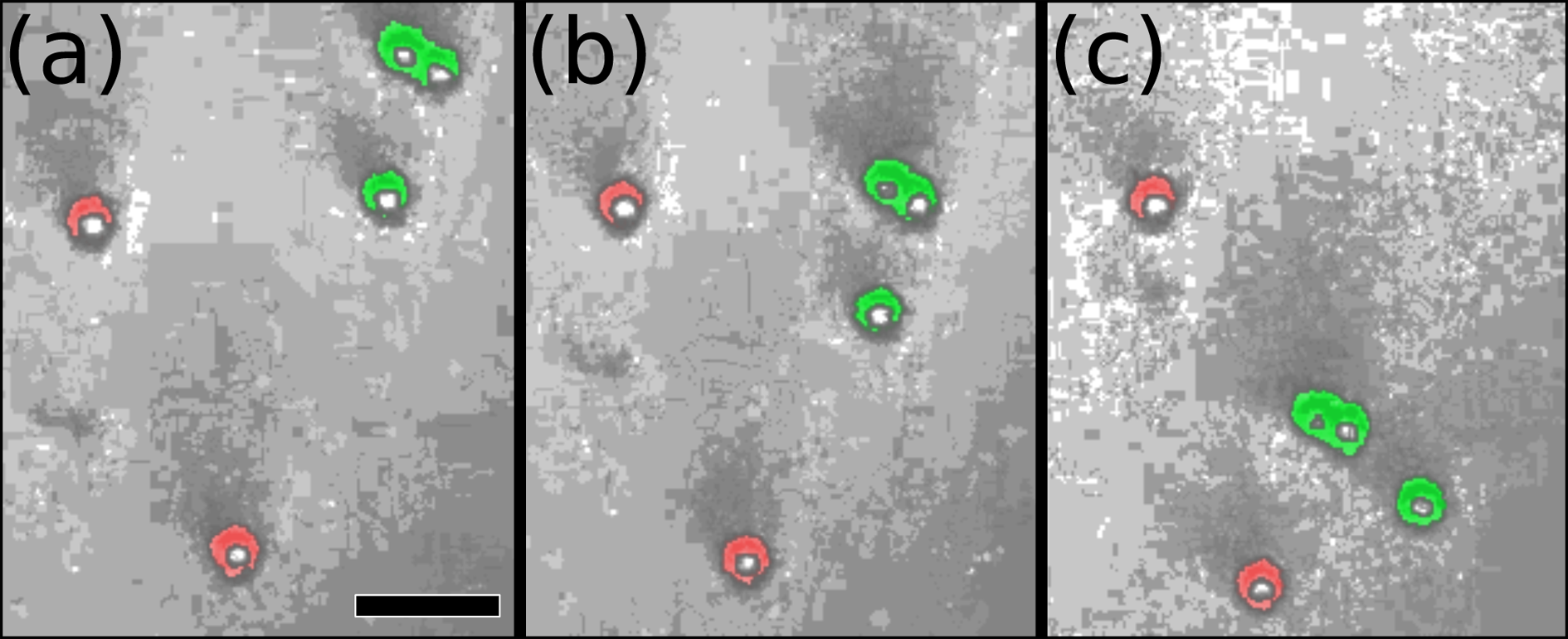}
\caption{Time sequence of microscope images showing a doublet (green) moving a singlet (green) out
	of its way. The other single colloids (red) remain at rest. Scale bar is $10$ $\mu$m.
	The images correspond to the times (a) $t=0$ s, (b) $t=2$ s, and (c)
        $t=4$ s. A video clip of the event recorded at twenty frames per second is provided in the supplemental
	material (adfig5.mp4).}
\label{fig5}
\end{figure}

Remarkably, none of the trains ever derails. This is due to the special properties of the confining
colloidal potential which are inherited from the square pattern.
In Fig.~\ref{fig5} we show the colloidal motion on a glass slide with no magnetic confining pattern.
Doublets also move when performing a modulation loop without the magnetic pattern (albeit 
not by a unit vector) and singlets are generically at rest. However due to the absence of the confining potential,
when one doublet moves onto a singlet, the singlet does not stay on track but is pushed to the side to let the doublet pass. 

\begin{figure}
	\includegraphics[width=1.\columnwidth]{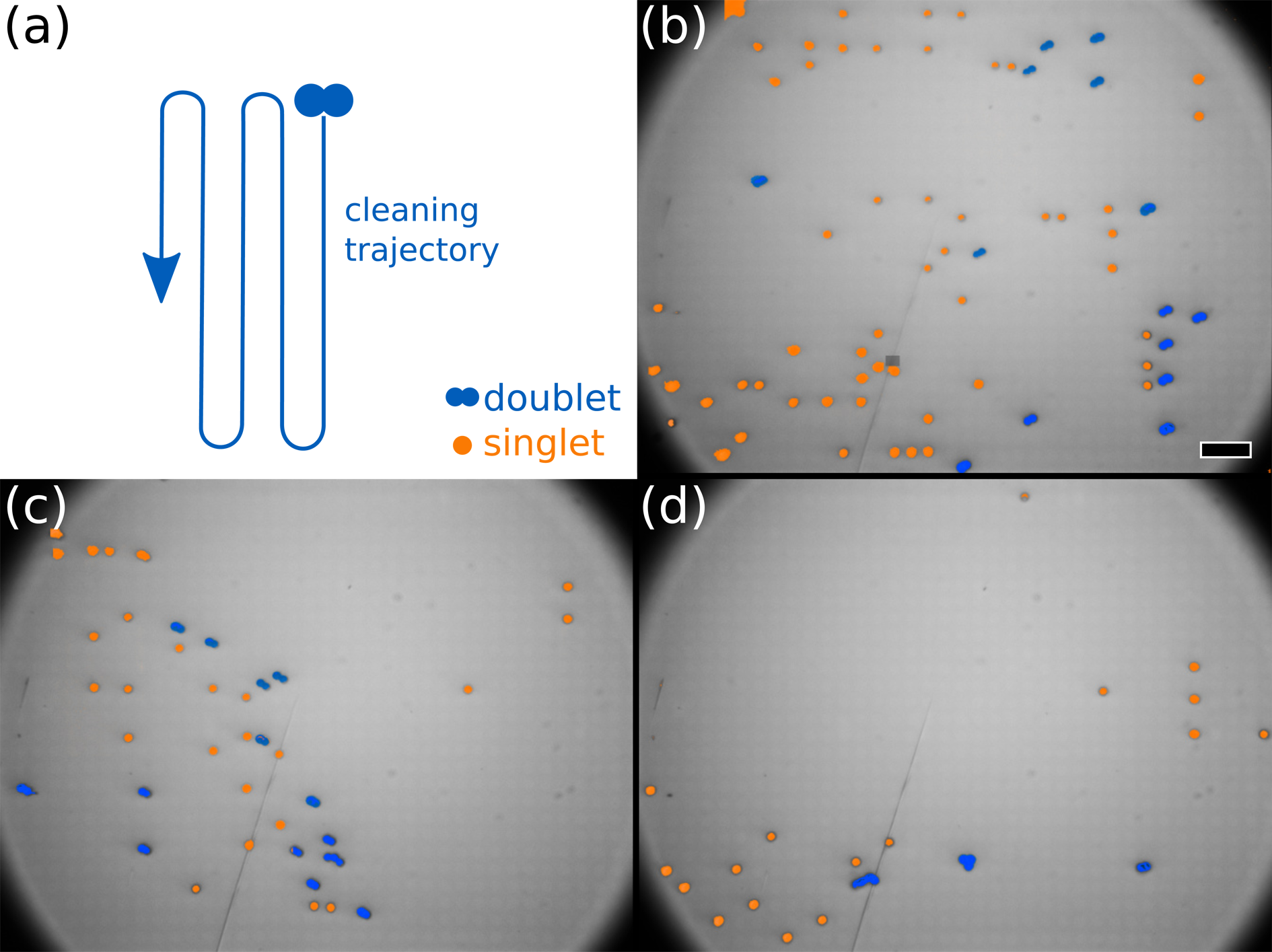}
	\caption{Surface cleaning. (a) Schematic of the trajectory of a doublet. Sequence of microscope
	images showing the cleaning of an area from single colloids (orange) by meandering doublets (blue)
	taken at times $t=0$ (b), $t=2$ min (c), and $t=6$ min (d). 
	The scale bar is $20$ $\mu$m. A video clip of the event is provided in the supplemental
		material (adfig6.mp4). }
	\label{fig6}
\end{figure}

So far we have shown the coordinated motion of colloids in one direction. However, by changing the global
orientation of the driving loop (winding around other fence points in $\cal C$) we can force the doublet
to move along any of the four symmetry directions of the magnetic pattern. Hence, the doublet-induced
motion of single colloids can potentially be used to arbitrarily set the position of the singlets across
the pattern. A step in this direction is shown in Fig.~\ref{fig6} where a complex modulation loop is programmed
to clean the surface of singlets.

Our colloidal trains are immersed into a low Reynolds number liquid where the propulsion of shape
changing objects is governed by the area enclosed by the loop in
shape space of the object \cite{Wilczek1,Wilczek2}. Swimmers are able to move by changing their shape. In contrast our 
biomimetic colloidal trains are driven by the shape of the potential created by the pattern and the external field which creates the topological
nature of this classical non-adiabatic phenomenon. We have shown in references~\cite{tp1,tp2,tp3,Rossi,Rossi2,Massana,CTI}
that, like other classical topological transport phenomena~\cite{Rechtsman,Mao,Murugan,Kane,Paulose,Nash,Huber},
there exist similarities with quantum mechanical topological transport~ \cite{Hasan,TI}. 

We have demonstrated how long range many-body interactions between the well separated colloidal particles can help sustain the
topological nature of the transport up to higher frequencies of driving. Such speeding up comes in
handy for lab-on-the-chip applications such as transporting loads from one place of a chip to another. 
Whether the doublet-induced motion of the singlets is caused by direct superadiabatic  non-equilibrium
interparticle interactions~\cite{PFT} or mediated by hydrodynamic interactions constitutes the subject of future studies.

%We don't need this in PRL. 
%\rem{\section*{Author contributions}{MMK, AdE, DdlH, \& TMF designed and performed the experiment, and wrote the manuscript with input from all the other authors.  MU \& FS produced the magnetic film. AT, RH, IK, ArE \& MR performed the fabrication of the micromagnetic domain patterns within the magnetic thin film. }}

%rem{\section*{Conflict of Interest}
%There are no conflicts to declare.}

\end{document}